\pdfoutput=1
\documentclass[twocolumn,fleqn,usenatbib]{mnras}

\usepackage{newtxtext,newtxmath}

\usepackage[T1]{fontenc}
\usepackage{ae,aecompl}

\usepackage{graphicx}	
\usepackage{amsmath}	
\usepackage{amssymb}	

\usepackage{longtable}
\usepackage{multirow}
\usepackage{lscape}

\usepackage{flushend}
\usepackage{scrextend}

\title[Statistical analysis of GGCs type properties]{Statistical analysis of Galactic globular cluster type properties}
\author[M.\,Simioni et al.]{
M.\,Simioni$^{1,2,3\thanks{email: msimioni@iac.es}}$, A. Aparicio$^{2,1}$, G. Piotto$^{3,4}$
\\
$^{1}$Instituto de Astrof{\'i}sica de Canarias, E-38200 La Laguna, Tenerife, Canary Islands, Spain\\
$^{2}$Department of Astrophysics, University of La Laguna, E-38200 La Laguna, Tenerife, Canary Islands, Spain\\
$^{3}$Dipartimento di Fisica e Astronomia ``Galileo Galilei'', Universit\`{a} degli Studi di Padova,  Vicolo dell'Osservatorio 3, Padova IT-35122\\
$^{4}$INAF - Osservatorio Astronomico di Padova, Vicolo dell'Osservatorio 5, I-35122 Padova, Italy\\
}
\date{Accepted 2020 March 24. Received 2020 March 15; in original form 2019 August 3}

\pubyear{2020}

\begin{document}
\label{firstpage}
\pagerange{\pageref{firstpage}--\pageref{lastpage}}
\maketitle

\begin{abstract}\label{sec:dyn:abstract}
The analysis of pseudo-colour diagrams, the so-called chromosome maps, of Galactic globular clusters (GCs) permits to classify them into type I and type II clusters. Type II GCs are characterized by an above-the-average complexity of their chromosome maps and some of them are known to display star-to-star variations of slow neutron-capture reaction elements including iron. This is at the basis of the hypothesis that type II GCs may have an extragalactic origin and were subsequently accreted by the Milky Way.

We performed a Principal Component Analysis to explore possible correlations among various GCs parameters in the light of this new classification. The analysis revealed that cluster type correlates mainly with relative age. The cause of this relation was further investigated finding that more metal-rich type II clusters, also appear to be younger and more distant from the Galactic centre. A depletion of type II clusters for positive values of Galactic coordinate Z was also observed, with no type II clusters detected above Z$\sim2$ kpc. Type II cluster orbits also have larger eccentricities than type I ones.
\end{abstract}



\defcitealias{2015AJ....149...91P}{Paper\,I}
\defcitealias{2017MNRAS.464.3636M}{Paper\,IX}
\defcitealias{2009ApJ...694.1498M}{MF09}

\section{Introduction}\label{sec:dyn:intro}
Seen as a single system, the Galactic Globular Clusters (GCs) were classically used to establish the non-preferential position of the Sun with respect to our Galaxy \citep{1918PASP...30...42S}. Being the oldest astronomical objects for wich reliable ages could be estimated, they were also used to put strong constraints to cosmological theories (see e.g. \citealt{2009ApJ...694.1498M}; \citealt{2016EAS....80..177C} and references therein).
In the Milky Way (MW) they populate a spheroid that extends out to various tens of kiloparsecs from the center of our Galaxy. A multitude of studies were oriented to them in order to find hints on MW formation processes (\citealt{1985ApJ...293..424Z}; {2001segc.book..223H}).

For the first time, \citet{1985ApJ...293..424Z} provided evidence that pointed to the existence of distinct sub-populations of GCs inside the MW. In his work, a halo population was detected in the Galaxy along with a disk population.
GCs associated to the halo population are characterized by lower metallicity and a small net rotation around the Galactic center. On the other hand, the disk population is more metal rich and has been found to have a velocity close to that of the local standard of rest.
At that time, this was used to promote the vision that, at least part of MW GCs have extragalactic origins and were later incorporated into MW via merging processes.

Since then, an increasing number of refinements in the number and components of distinct GC populations in the MW was proposed.
\citet{1995MNRAS.277..781I} discovered the Sagittarius dwarf galaxy. Its chemical and kinematical patterns were detected in a large area of the sky supporting the idea that this satellite of the MW is being subject to tidal stripping. The authors, also suggested that some MW GCs originally were part of this galaxy.
The existence of others nearby, stripped dwarf galaxies was proposed, along with a list of GCs likely associated with them (see for example \citealt{2009ApJ...694.1498M} --- hereafter \citetalias{2009ApJ...694.1498M} and references therein).

More recently, in the first decade of this century, the advent of deep, high resolution photometry of Galactic GCs, led to the detection of distinct and detached sequences at different stellar evolutionary stages in the Color-Magnitude Diagrams (CMD) of these objects. This result, coupled with the growing number of spectroscopical evidence, was interpreted in terms of multiple stellar populations (MPs, \citealt{2012A&ARv..20...50G}; \citealt{2018ARA&A..56...83B}).
It is worth mentioning that distinct sequences of stars, that appear detached in the CMD, put strong constraints to modelling the processes of star formation in GCs. This evidence, in fact, rules out a continous star formation process and suggests that bursts of star formation occurred in these systems (\citealt{2015AJ....149...91P} --- hereafter \citetalias{2015AJ....149...91P}, but see also  \citealt{2013MNRAS.436.2398B} and \citealt{2014ApJ...797...59H}).

At first, the MPs phenomenon was detected in the CMD of single GCs. But with the completion of a large photometric survey, namely the {\it Hubble Space Telescope} (HST) UV Legacy Survey of Galactic Globular Clusters (PI: Piotto, \citetalias{2015AJ....149...91P}), it has been confirmed that this phenomenon is a common characteristic of virtually all Galactic GCs (\citetalias{2015AJ....149...91P}; \citealt{2017MNRAS.464.3636M} --- hereafter \citetalias{2017MNRAS.464.3636M}).

Even more interesting is that, with the introduction of the chromosome maps \citepalias{2017MNRAS.464.3636M}, MW GC population has been once again subdivided.
These two-color diagrams, in fact, fully exploit the potential of the UV observations taken in the context of  the HST UV Legacy Survey of Galactic GCs. Specifically, they maximize the color difference between the various stellar populations inside each GC, optimizing their detection and characterization (\citetalias{2015AJ....149...91P}; \citetalias{2017MNRAS.464.3636M}).
Indeed, in \citetalias{2017MNRAS.464.3636M} it was shown that the chromosome maps of the majority of the observed GCs show the presence of two major distinct groups of stars (1G population and 2G population). Samples of stars taken from these two groups display chemical abundance differences in light elements; for example the well known Na-O anticorrelation.
These clusters were named type I clusters. 

The remaining GCs show more complex chromosome maps: other groups of stars can be detected, in addition to the two main groups present in type I clusters, all with their own Na-O anticorrelation. These additional groups of stars also have different Ba and Fe abundances, and more in general a different abundance of elements produced via slow neutron capture reactions (\citetalias{2017MNRAS.464.3636M}). In this respect, it is worth mentioning that the presence of small intrinsic iron spread, at least for some GCs, is still debated, as they can be artificially introduced by the method used to derive atmospheric parameters of stars (\citealt{2015ApJ...801...69M}, \citealt{2016MNRAS.457...51L}; but see also \citealt{2016ApJS..226...16L}).
GCs that have chromosome maps of this kind are defined as type II clusters. In \citetalias{2017MNRAS.464.3636M}, 10 type II clusters have been detected: NGC\,362, NGC\,1261, NGC\,1851, NGC\,5139, NGC\,5286, NGC\,6388, NGC\,6656, NGC\,6715, NGC\,6934 and NGC\,7089. Subsequently, by means of a detailed characterization of the CM of M15, \citet{2018MNRAS.477.2004N} have re-classified this cluster as type II. We thus added NGC\,7078 to the list of type II GCs.

In this work we performed a principal component analysis (PCA) in order to investigate the correlation of cluster type with other GC parameters.
Noteworthy previous applications of this technique to GCs is by \citet{1994AJ....108.1292D}, who identified a correlation between cluster luminosities and concentration, and by \citet{2006A&A...452..875R}, who examined the effect of various parameters on the morphology of the horizontal branch revealing the influence of cluster mass, thus providing hints of self-enrichment in Galactic globular clusters. Another example is \citet{2010A&A...516A..55C} who studied the effect of detailed chemical composition of the distinct stellar populations in GCs.

The structure of this paper is as follows: in Section \ref{sec:dyn:pca} we present the details of the PCA and briefly discuss the results. In Section \ref{sec:dyn:rela} possible relations among various cluster parameters are presented in more details. In particular, we reviewed the age-metallicity and age-Galactocentric distance relations (Section \ref{sec:dyn:agemet}); the metallicity vs. Galactocentric distance relation (Section \ref{sec:dyn:distmet}); the relation between orbital parameters and cluster type (Section \ref{sec:dyn:orb}); the mass distribution (Section \ref{sec:dyn:mass}); and finally the spatial distribution of GCs (Section \ref{sec:dyn:spdist}).

\section[Variables of the problem and PCA]{Variables of the problem and Principal Component Analysis}\label{sec:dyn:pca}

We investigated a large sample of variables through principal component analysis. In particular, we focus on $11$ quantities, reported in Table \ref{tab:dyn:pcain}: cluster type, Oosterhoff type, metallicity, relative age, orbit total energy, orbit total angular momentum, orbit eccentricity, orbit inclination with respect to the Glactic plane, cluster masses, core radii and tidal radii.
This sample of GC properties was chosen in order to have a complete view of the relations affecting the cluster type.
In particular, these quantities were collected from several catalogues that provide information for different GC samples. 
Our final sample consist on 25 GCs, 7 of them of type II and the rest of type I. The 7 type II clusters are NGC\,362, NGC\,1851, NGC\,5139, NGC\,6656, NGC\,6934, NGC\,7078, NGC\,7089. The full list of parameters of our sample is given in Table \ref{tab:dyn:pcasmpl}.

\begin{table}
  \centering
  \caption{GC properties considered for the PCA. They will be identified in the following with the identification number provided in the first column. In the third column we report the sources from where values were taken. References are: i, \protect\citetalias{2017MNRAS.464.3636M}; ii, \protect\citet{2009Ap&SS.320..261C}; iii, \protect\citet{2010AJ....139..357Z}; iv, \protect\citet{2010MNRAS.406..329S}; v, \protect\citet{2011PASP..123.1044V}; vi, \protect\citet{2013AJ....146..119K}; vii, \protect\citet{2009A&A...508..695C}; viii, \protect\citetalias{2009ApJ...694.1498M}; ix, \protect\citet{1999AJ....117.1792D}; x, \protect\citet{1997ApJ...474..223G}; xi, \protect\citet{1996AJ....112.1487H}.}
  \label{tab:dyn:pcain}
  \begin{tabular}{rll}
      & Description               & References         \\
    \hline
    1 & Cluster type              & i                  \\
    2 & Oosterhoff type           & ii; iii; iv; v; vi \\
    3 & Metallicity               & vii                \\
    4 & Relative age              & viii               \\
    5 & Total energy of the orbit & ix                 \\
    6 & Total angular momentum    & ix                 \\
    7 & Orbit eccentricity        & ix                 \\
    8 & Orbit inclination with    & ix                 \\
      & respect to the Gal. plane &                    \\
    9 & Cluster mass              & x                  \\
   10 & Core radius (parsec)      & xi                 \\
   11 & Tidal radius (parsec)     & xi                 \\
  \end{tabular}
\end{table}
  \begin{table*}
  \centering
  \caption{The considered sample of Galactic GC. For each cluster the values of the parameters used in the PCA are given. Labels in the first row identify the parameters, according to the designations given in Table \ref{tab:dyn:pcain}. They are respectively: cluster type (1), Oosterhoff type (2), metallicity (3), relative age (4), total energy of the orbit (5), total angular momentum (6), orbit eccentricity (7), orbit inclination with respect to the Galactic plane (8), cluster mass (9), core radius (10), tidal radius (11).}
  \label{tab:dyn:pcasmpl}
  \begin{tabular}{cccccccccccc}
   Parameter: & 1 & 2 & 3 & 4 & 5 & 6 & 7 & 8 & 9 & 10 & 11 \\
              &   &   &   &   & $(10^{2}\,{\rm km}^{2}\,{\rm s}^{-2})$ & $({\rm kpc}\,{\rm km}\,{\rm s}^{-1})$ &  & (deg) & $({\rm M}_{\odot})$ & (pc) & (pc) \\  
    \hline    
    NGC\,104  & I  & II  & -0.76 & 1.02 &  -872 & 1274.0 & 0.17 & 29 & 1.45e+06 & 0.4712 & 55.36 \\
    NGC\,288  & I  & II  & -1.32 & 0.83 &  -787 &  654.1 & 0.74 & 44 & 1.11e+05 & 3.4950 & 34.15 \\
    NGC\,362  & II & I   &  -1.3 & 0.81 &  -856 &  358.0 & 0.85 & 21 & 3.78e+05 & 0.4503 & 25.91 \\
    NGC\,1851 & II & I   & -1.18 & 0.78 &  -340 & 2259.0 & 0.69 & 22 & 5.61e+05 & 0.3168 & 22.95 \\
    NGC\,2298 & I  & II  & -1.96 & 0.99 &  -655 & 1394.0 & 0.78 & 36 & 7.20e+04 & 0.9739 & 23.36 \\
    NGC\,4590 & I  & II  & -2.27 &  0.9 &  -396 & 3160.0 & 0.48 & 30 & 3.06e+05 & 1.7380 & 44.67 \\
    NGC\,5024 & I  & II  & -2.06 & 0.99 &  -203 & 4614.0 & 0.40 & 62 & 7.33e+05 & 1.8220 & 95.64 \\
    NGC\,5139 & II & Int & -1.64 &  0.9 & -1089 &  417.3 & 0.67 & 16 & 2.64e+06 & 3.5850 & 73.19 \\
    NGC\,5272 & I  & I   &  -1.5 & 0.89 &  -649 & 1531.0 & 0.42 & 55 & 7.82e+05 & 1.0980 & 85.22 \\
    NGC\,5466 & I  & II  & -2.31 & 1.06 &   -83 & 2900.0 & 0.79 & 42 & 1.33e+05 & 6.6560 & 72.98 \\
    NGC\,5904 & I  & I   & -1.33 & 0.83 &  -289 & 1344.0 & 0.87 & 33 & 8.34e+05 & 0.9599 & 51.55 \\
    NGC\,6093 & I  & II  & -1.75 & 0.98 & -1311 &  374.1 & 0.73 & 39 & 3.67e+05 & 0.4363 & 20.88 \\
    NGC\,6121 & I  & I   & -1.18 & 0.98 & -1121 &  142.4 & 0.80 & 23 & 2.25e+05 & 0.7423 & 33.16 \\
    NGC\,6171 & I  & I   & -1.03 & 1.09 & -1198 &  519.5 & 0.21 & 44 & 2.04e+05 & 1.0430 & 35.33 \\
    NGC\,6205 & I  & II  & -1.58 & 0.91 &  -476 & 2156.0 & 0.62 & 54 & 6.27e+05 & 1.2800 & 43.39 \\
    NGC\,6341 & I  & II  & -2.35 & 1.03 &  -880 &  337.8 & 0.76 & 23 & 3.64e+05 & 0.6277 & 30.05 \\
    NGC\,6362 & I  & I   & -1.07 & 1.06 & -1090 &  655.6 & 0.39 & 22 & 1.17e+05 & 2.4980 & 30.73 \\
    NGC\,6584 & I  & I   &  -1.5 & 0.88 &  -773 &  573.5 & 0.87 & 20 & 2.19e+05 & 1.0210 & 30.13 \\
    NGC\,6656 & II & II  &  -1.7 & 0.99 &  -871 & 1008.0 & 0.53 & 18 & 5.36e+05 & 1.2380 & 29.70 \\
    NGC\,6779 & I  & II  &    -2 & 1.07 &  -791 &  392.9 & 0.86 & 15 & 1.89e+05 & 1.2030 & 28.86 \\
    NGC\,6809 & I  & II  & -1.93 & 0.96 & -1038 &  763.1 & 0.51 & 56 & 2.41e+05 & 2.8270 & 24.07 \\
    NGC\,6934 & II & I   & -1.56 & 0.87 &  -249 & 2905.0 & 0.72 & 55 & 2.15e+05 & 0.9983 & 33.83 \\
    NGC\,7078 & II & II  & -2.33 & 1.01 &  -752 & 1631.0 & 0.32 & 36 & 9.84e+05 & 0.4235 & 82.58 \\
    NGC\,7089 & II & II  & -1.66 & 0.92 &  -290 & 2384.0 & 0.68 & 53 & 8.81e+05 & 1.0700 & 41.65 \\
    NGC\,7099 & I  & II  & -2.33 & 1.01 &  -937 &  820.4 & 0.39 & 52 & 2.74e+05 & 0.1414 & 44.71 \\
  \end{tabular}
  \end{table*}
\begin{figure*}
  \includegraphics[width=\textwidth]{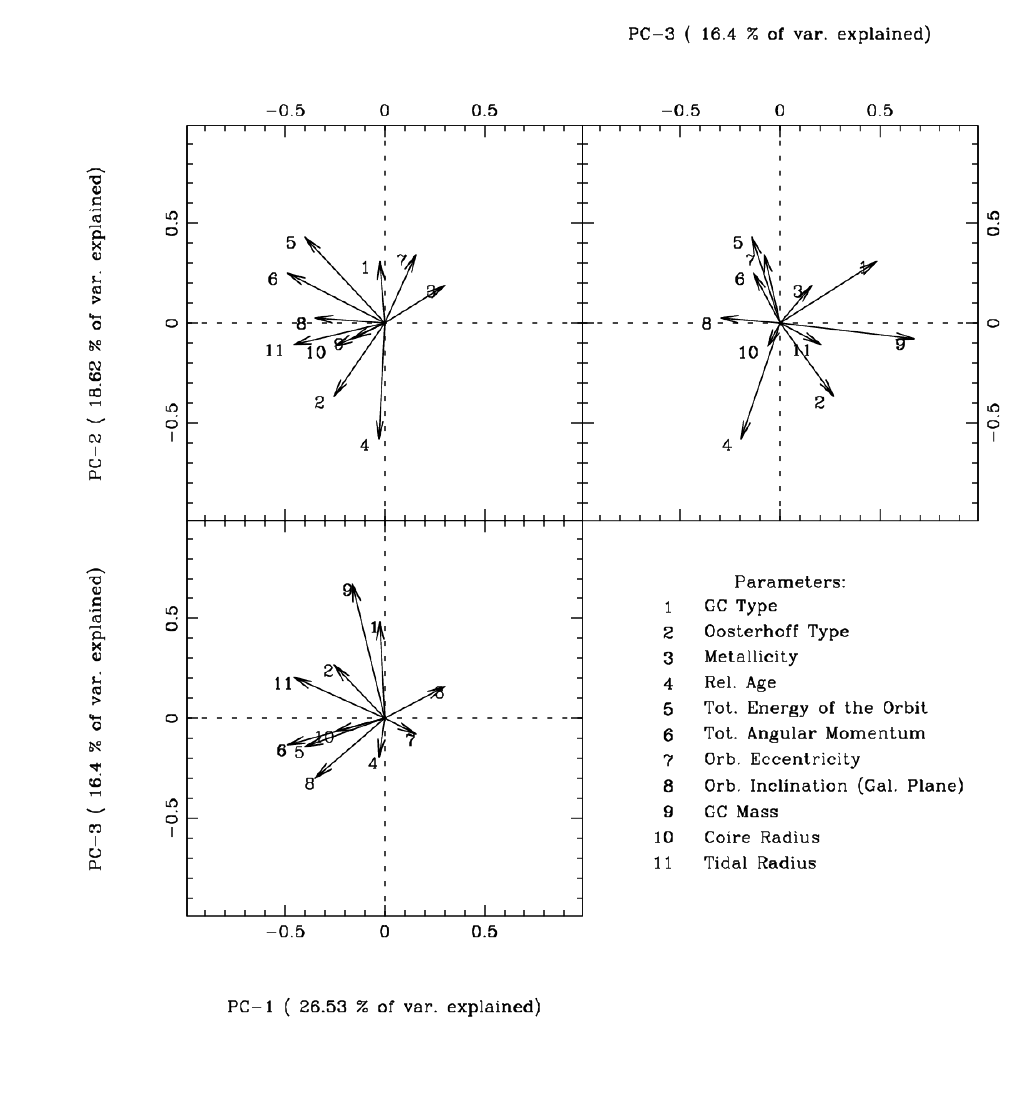}
  \caption{Eigenvectors projections of the 11 GC parameters on the planes defined by the first three principal components. Each axis legend reports also the total fraction of variance explained by each component. See text for details.}
  \label{fig:dyn:pca123}
\end{figure*}
\begin{table}
  \centering
  \small
  \caption{For each principal component, we report the corresponding eigenvalues in the second column, the resulting fraction of the total variance explained in the third column and, in the last column, the cumulative fraction of total variance explained.}
  \label{tab:dyn:pcaval}
  \begin{tabular}{cccc}
        & Eigenvalue & Frac. of tot. var. & Cumul. frac. of tot. var. \\
  \hline
  PC-1  & $ 2.919$ & $26.5\%$ & $ 26.5\%$ \\
  PC-2  & $ 2.049$ & $18.6\%$ & $ 45.1\%$ \\
  PC-3  & $ 1.804$ & $16.4\%$ & $ 61.5\%$ \\
  PC-4  & $  1.49$ & $13.5\%$ & $ 75.1\%$ \\
  PC-5  & $0.9198$ & $ 8.4\%$ & $ 83.5\%$ \\
  PC-6  & $0.6084$ & $ 5.5\%$ & $ 89.0\%$ \\
  PC-7  & $0.5202$ & $ 4.7\%$ & $ 93.7\%$ \\
  PC-8  & $0.3541$ & $ 3.2\%$ & $ 96.9\%$ \\
  PC-9  & $0.2161$ & $ 2.0\%$ & $ 98.9\%$ \\
  PC-10 & $ 0.069$ & $ 0.6\%$ & $ 99.5\%$ \\
  PC-11 & $0.0509$ & $ 0.5\%$ & $100.0\%$ \\
  \end{tabular}
\end{table}
  \begin{table*}
    \centering
    \caption{Correlations between principal components (columns) and GC parameters (rows).}
    \begin{tabular}{c|ccccccccccc}
      & PC-1    & PC-2    & PC-3    & PC-4    & PC-5    & PC-6    & PC-7    & PC-8    & PC-9    & PC-10   & PC-11   \\
      \hline
       1 & $-0.0447$ & $ 0.4403$ & $ 0.6521$ & $-0.0620$ & $-0.3411$ & $-0.2166$ & $ 0.4174$ & $-0.1817$ & $-0.0602$ & $-0.0068$ & $ 0.0296$ \\
       2 & $-0.4369$ & $-0.5260$ & $ 0.3584$ & $-0.4643$ & $-0.1679$ & $ 0.2595$ & $ 0.0819$ & $ 0.2742$ & $-0.0043$ & $-0.0034$ & $ 0.1051$ \\
       3 & $ 0.5147$ & $ 0.2687$ & $ 0.2136$ & $ 0.5317$ & $ 0.4960$ & $ 0.0178$ & $ 0.1267$ & $ 0.2081$ & $-0.1472$ & $-0.0250$ & $ 0.0809$ \\
       4 & $-0.0500$ & $-0.8305$ & $-0.2620$ & $ 0.0387$ & $-0.0758$ & $-0.3879$ & $ 0.0620$ & $ 0.0179$ & $-0.2774$ & $ 0.0033$ & $-0.0165$ \\
       5 & $-0.6836$ & $ 0.6176$ & $-0.1894$ & $-0.0993$ & $ 0.0175$ & $-0.1861$ & $-0.0923$ & $ 0.1583$ & $-0.0764$ & $ 0.1755$ & $ 0.0216$ \\
       6 & $-0.8364$ & $ 0.3562$ & $-0.1766$ & $ 0.1437$ & $-0.0804$ & $-0.1713$ & $ 0.0298$ & $ 0.2313$ & $ 0.0279$ & $-0.1703$ & $-0.0422$ \\
       7 & $ 0.2660$ & $ 0.4923$ & $-0.1068$ & $-0.7404$ & $ 0.0924$ & $ 0.1035$ & $-0.1849$ & $-0.0871$ & $-0.2449$ & $-0.0789$ & $ 0.0084$ \\
       8 & $-0.5949$ & $ 0.0371$ & $-0.4005$ & $ 0.3704$ & $-0.0926$ & $ 0.4913$ & $ 0.2178$ & $-0.1432$ & $-0.1694$ & $ 0.0136$ & $-0.0180$ \\
       9 & $-0.2736$ & $-0.1099$ & $ 0.9025$ & $ 0.0843$ & $ 0.1310$ & $ 0.1190$ & $-0.1621$ & $ 0.0637$ & $-0.1077$ & $ 0.0253$ & $-0.1326$ \\
      10 & $-0.4286$ & $-0.1697$ & $-0.0860$ & $-0.4580$ & $ 0.6624$ & $-0.0575$ & $ 0.3363$ & $-0.0827$ & $ 0.0832$ & $ 0.0088$ & $-0.0362$ \\
      11 & $-0.7801$ & $-0.1518$ & $ 0.2747$ & $ 0.2320$ & $ 0.2088$ & $-0.0991$ & $-0.2983$ & $-0.2911$ & $ 0.0013$ & $-0.0366$ & $ 0.1031$ \\
    \end{tabular}
  \label{tab:dyn:corr}
  \end{table*}

The resulting principal component values, along with the fraction of total variance explained by each component are listed in Table \ref{tab:dyn:pcaval}.
The first four components explain respectively $26\%$, $19\%$, $16\%$ and $13\%$ of the total variance of the sample, for a total of $74\%$.
Table \ref{tab:dyn:corr} lists the correlation coefficients between principal components and the 11 quantities considered in the analysis.

Figure \ref{fig:dyn:pca123} shows the eigenvector projections of the 11 quantities considered in the planes defined by the first and the second components (upper-left panel), the first and the third components (lower panel) and the third and the second components (upper-right panel).

The first component correlates mainly with orbital parameters like total energy, total angular momentum and inclination (5, 6 and 8 respectively); it also correlates with tidal radius (11), which points to a connection between this parameter and orbital ones. To a lesser extent it also correlates with Oosterhoff type (2) and metallicity (3), confirming the well known relation between metallicity and Oosterhoff type.

The second component shows correlations with relative age (4), Oosterhoff type (2), orbit total energy (5) and eccentricity (7). It can be also observed a mild correlation between this component and cluster type (4). In particular, the alignment of the eigenvectors associated to cluster type and relative age in the upper-left panel of Figure \ref{fig:dyn:pca123}, call for further investigation on its origin. We will discuss this result in the following section.

The third component mainly correlates with GC mass (9) and cluster type (4) pointing to a correlation between both parameters. 

The fourth component mainly correlates with the orbit eccentricity and with the metallicity. This relation was discussed in \citet{1999AJ....117.1792D}. This component mildly also correlates with Oosterhoff type and core radius (2 and 10 respectively) both in the same sense as eccentricity.

Since the first component mainly correlates only with orbital parameters and tidal radius, the conclusion that can be reached is that it describes general properties of the whole GC sample. But our interest is to investigate how the new classification in type I and type II clusters affects the sample, so we concentrated on the second and third components, which have the strongest correlation with cluster type. The relations between it and other parameters are shown in the upper-right panel of Figure \ref{fig:dyn:pca123}. A possible relation between cluster type (1) and relative age (4) can be observed. In fact, the two associated vectors points in opposite directions and are almost parallels.

\section{Relations studied}\label{sec:dyn:rela}
The PCA analysis performed in the previous section highlighted some possible correlations between various GC properties. In particular, we are interested in studying those that involve cluster type. For this reason, we reviewed in more details some of them, this time without the use of PCA, but quantifying the statistical significance of each one. Our choice is also motivated by the fact that the sample of clusters we could use in PCA is more limited than the entire sample used in this paper because of the lack of some of the parameters used in the PCA.

For clarity, we report in Table \ref{tab:dyn:sprkc} the Spearman rank coefficients (${\rm r_{S}}$) of the various correlations we investigated. In the first column we report the coefficients for type I clusters. The coefficients for type II ones are in the second column and, in the last column, we provide ${\rm r_{S}}$ for the sample of type II clusters excluding NGC\,6388. This cluster is among the type II clusters that are not present in the sample used for the PCA in Section \ref{sec:dyn:pca} beacuse we could not find some of the parameters used in the analysis. Its metallicity and position in the Galaxy make it rather peculiar among this group of GCs (see also \citetalias{2017MNRAS.464.3636M}). 
          
\subsection{Age-metallicity relation} \label{sec:dyn:agemet}
The strong anti-correlation found between cluster type and age deserves further investigation. We thus considered, for this purpose, the sample of relative ages and metallicities of \citetalias{2009ApJ...694.1498M}.

\begin{figure*}
  \includegraphics[width=\textwidth]{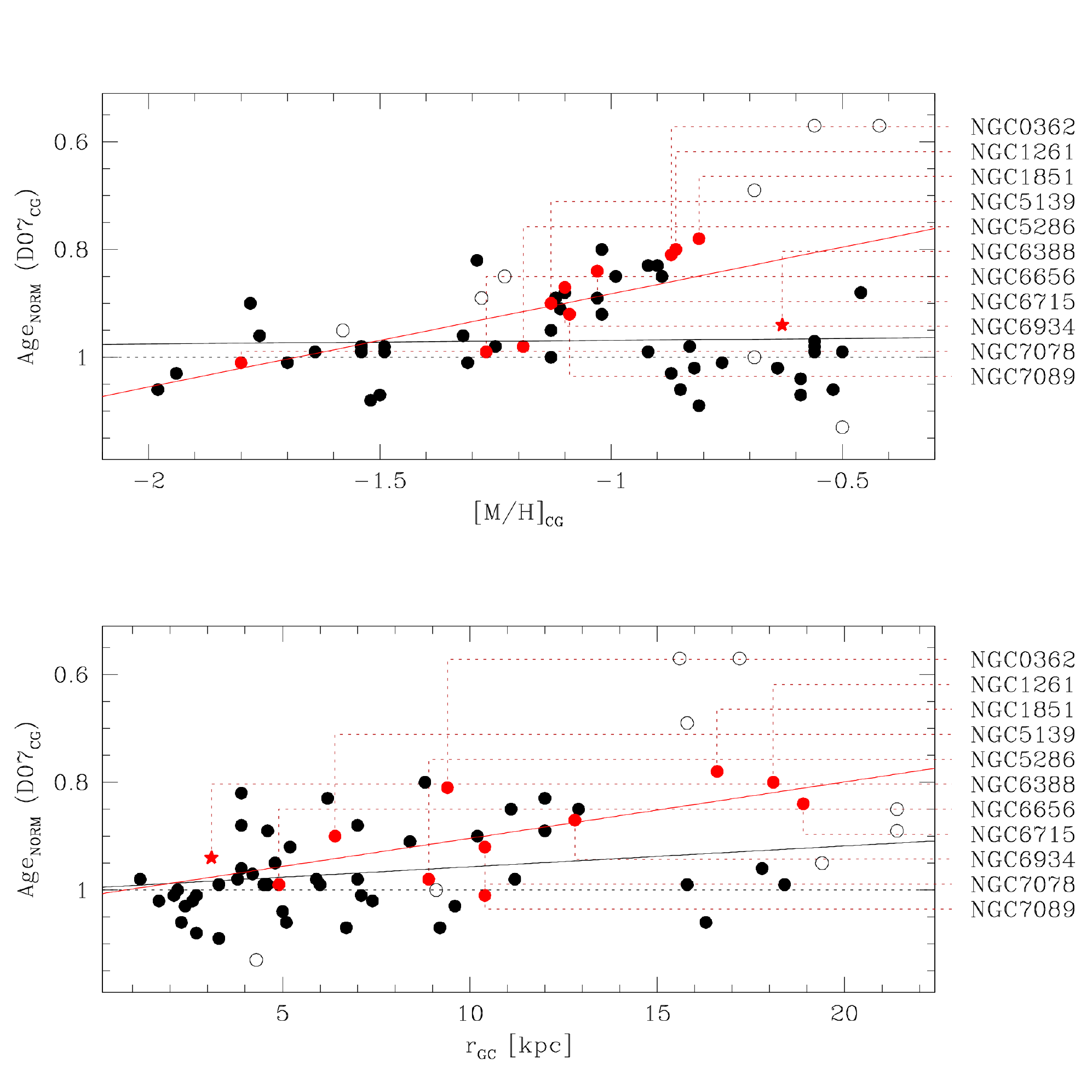}
  \caption{Age-metallicity relation for Galactic GCs from Table 1 of \citetalias{2009ApJ...694.1498M}. The displayed metallicities are in the \citet{1997A&AS..121...95C} scale, while distances from Galactic center come from \citet{1996AJ....112.1487H}. Empty circles represent GCs for which cluster type has not been determined in \citetalias{2017MNRAS.464.3636M}, black points type I clusters and red dots type II ones. The red star is associated to NGC\,6388. In both panels, the regression lines are shown in black and red for type I and type II GCs respectively.}
  \label{fig:dyn:spv}
\end{figure*}

One of the main results coming from \citetalias{2009ApJ...694.1498M} is that the old population of globular clusters shows a small relative age dispersion of about 0.05, with no detectable trends of age with metallicity. On the other hand, young clusters show a trend with metallicity, the more metal rich clusters being younger.
This trend can be clearly seen in Figure 10 of \citetalias{2009ApJ...694.1498M}, that we have reproduced in Figure \ref{fig:dyn:spv}, highlighting type I and type II clusters present in their sample with filled black dots and filled red symbols, respectively. In particular, we marked the position of NGC\,6388 with a red star. Of the total $64$ GCs considered in \citetalias{2009ApJ...694.1498M}, fortyfive of the $64$ GCs in the \citetalias{2009ApJ...694.1498M} sample are of Type I while $11$ are of Type II.   
Interestingly enough, all type II clusters except NGC\,6388, show a clear age-metallicity relation.

\citetalias{2009ApJ...694.1498M} noted that the age-metallicity trend of young clusters is followed by GCs believed to be associated with accreted dwarf galaxies, namely Sagittarius (\citealt{1995MNRAS.277..781I}; \citealt{2000AJ....120.1892D}; \citealt{2003AJ....125..188B}), Monoceros (\citealt{2003ApJ...594L.119C}; \citealt{2004ApJ...602L..21F}) and the stellar overdensity in Canis Major \citep{2004MNRAS.355L..33M}.
The cluster sample of possible extra-galactic origin was composed by: Terzan\,7, Terzan\,8,Arp\,2, NGC\,6715 (M54), Pal\,12 and NGC\,4147 (associated to Sagittarius dwarf); NGC\,2298, NGC\,2808, NGC\,5286, Pal\,1 and BH\,176 (Monoceros stream).
NGC\,2298 and NGC\,2808 are also present in the list of GCs presumably associated to the Canis Major stellar overdensity along with NGC\,1851 and NGC\,1904.
Three of them (NGC\,6715, NGC\,5286 and NGC\,1851) are type II clusters, while NGC\,2298 and NGC\,2808 result to be type I clusters. Thus, not all GCs in the young group of \citetalias{2009ApJ...694.1498M} are type II.

Figure \ref{fig:dyn:spv} shows that NGC\,6388 is the only type II cluster that does not follow the age-metallicity trend traced by the others. It may be worth noting that NGC\,6388 is also the type II cluster which has the smallest distance to the Galactic center and also the one with the highest metallicity. It is the only type II cluster that can be strictly ascribed to the bulge population.

\begin{table}
  \centering
  \caption{Spearman rank coefficients for different relations studied. For each case we give the coefficients for the type I clusters (TI), for the whole type II cluster sample (TII) and, in the last column, for the sample of type II cluster excluding NGC\,6388 from it.}
  \begin{tabular}{lccc}
  Relation                                                         & TI    & TII   & halo TII \\
  \hline                                                              
  Relative Age vs. ${\rm [Fe/H]}$                                  &  0.08 & -0.72 & -0.96    \\
  Relative Age vs. ${\rm r_{GC}}$                                   & -0.32 & -0.64 & -0.64   \\
  Relative Age vs. ${\rm \mid X-8 \mid}$                           & -0.31 & -0.63 & -0.65    \\
  Relative Age vs. ${\rm \mid Y \mid}$                             & -0.15 & -0.40 & -0.34    \\
  Relative Age vs. ${\rm \mid Z \mid}$                             & -0.18 & -0.75 & -0.74    \\
  ${\rm [Fe/H]}$ vs. ${\rm r_{GC}}$ (${\rm [Fe/H]}\leq -1$)         & -0.39 &  0.61 &  0.61   \\
  ${\rm [Fe/H]}$ vs. ${\rm \mid X-8 \mid}$ (${\rm [Fe/H]}\leq -1$) & -0.13 &  0.70 &  0.70    \\
  ${\rm [Fe/H]}$ vs. ${\rm \mid Y \mid}$ (${\rm [Fe/H]}\leq -1$)   & -0.33 &  0.27 &  0.27    \\
  ${\rm [Fe/H]}$ vs. ${\rm \mid Z \mid}$ (${\rm [Fe/H]}\leq -1$)   & -0.27 &  0.72 &  0.72    \\
  \end{tabular}
  \label{tab:dyn:sprkc}
\end{table}

Apart from NGC\,6388, all the other type II clusters seem to follow a well defined trend. In particular, for the age-metallicity relation, the ${\rm r_{S}}$ calculated for all type II clusters is $-0.71$. The probability of randomly extracting eleven points from the sample of \citetalias{2017MNRAS.464.3636M} with the same value of ${\rm r_{S}}$ or higher (i.e. ${\rm r_{S}}\leq -0.71$), is $\sim 4.5\%$. If NGC\,6388 is removed from the sample of type II clusters, ${\rm r_{S}}=-0.96$ , and the probability of randomly extracting $10$ clusters with ${\rm r_{S}}\leq-0.96$ results less than $0.1\%$.

On the other hand, the small ${\rm r_{S}}$ calculated for type I clusters ($0.08$) is consistent with the hypothesis that all GCs of the old population (as defined by \citetalias{2009ApJ...694.1498M}) are preferentially Type I clusters. With a mean relative age of $0.97$ (median $0.99$ and standard deviation of $0.08$) they are slightly biased towards older ages than type II GCs which result to have a mean relative age of $0.89$ and standard deviation of $0.08$. This result possibly explain the relation between age and cluster type highlighted by the PCA.

The lower panel of Figure \ref{fig:dyn:spv} is even more interesting. Here we plot the relative age against the Galactocentric distance (${\rm r_{GC}}$). \citetalias{2009ApJ...694.1498M} noted an increase in age dispersion with distance for the young group of GCs. Figure \ref{fig:dyn:spv} suggests that this trend may be, at least in part, the consequence of the previously observed age-metallicity relations for these GCs.
Focussing the attention on type II GCs (red dots), what Figure \ref{fig:dyn:spv} shows is an age-${\rm r_{GC}}$ relation for these clusters, rather than an increase in age spread. We calculated ${\rm r_{S}}$ for both type I and II clusters in the plane defined by age and ${\rm r_{GC}}$ and found ${\rm r_{S}}=-0.32$ for type I clusters and ${\rm r_{S}}=-0.64$ for type II ones (see Table \ref{tab:dyn:sprkc}).
The probability of obtaining the same, or higher correlation (${\rm r_{S}}\leq-0.64$) randomly extracting eleven points from the sample of type I and type II clusters of \citetalias{2009ApJ...694.1498M} is $\sim 17\%$.

To further analyze the possible age-${\rm r_{GC}}$ relation found for type II clusters we also considered each single cartesian Galactocentric coordinate (X, Y, Z) defined in the ususal way and measured in kiloparsec. We studied the correlation between age and ${\rm \mid X-8 \mid}$, ${\rm \mid Y \mid}$ and ${\rm \mid Z \mid}$ respectively for each sample, in analogy to what has been done before. We report the results in Table \ref{tab:dyn:sprkc}.
${\rm r_{S}}$ for type II clusters: ${\rm r_{S}}=-0.63$ for X, ${\rm r_{S}}=-0.40$ for Y and ${\rm r_{S}}=-0.75$ for Z. It thus appears that the possible age-${\rm r_{GC}}$ relation for type II clusters is more significant for the Z Galactic coordinate.

\subsection{$[{\rm Fe}/{\rm H}]$ vs Galactocentric distance for the Type I and Type II clusters} \label{sec:dyn:distmet}

In this section we explore the distribution of values in the metallicity-${\rm r_{GC}}$ plane.

\begin{figure*}
  \includegraphics[width=\textwidth]{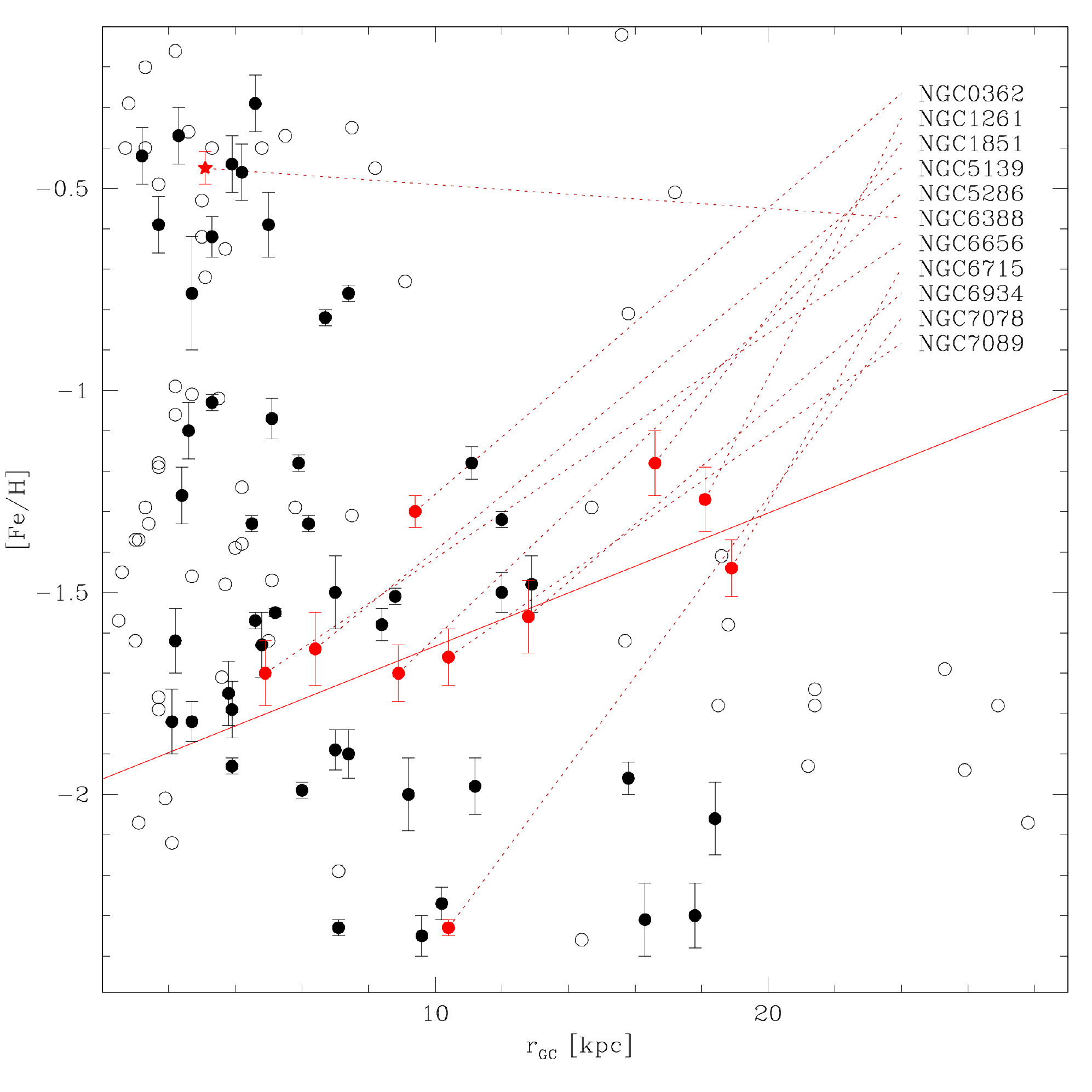}
  \caption{Galactocentric distances (from \citealt{1996AJ....112.1487H}) versus metallicities (from \citealt{2009A&A...508..695C}). Empty dots represent cluster not included in this survey, black dots refer to type I GCs and red dots to type II. NGC\,6388 position is marked with a red star. The red solid line is the resulting regression line for type II clusters with ${\rm [Fe/H]}\leq -1$, excluding NGC\,6388.}
  \label{fig:dyn:distmetal}
\end{figure*}

In Figure \ref{fig:dyn:distmetal} the Galactocentric distances from \citealt{1996AJ....112.1487H} against the $[{\rm Fe}/{\rm H}]$ values from \citet{2009A&A...508..695C} are plotted. The distribution of type I clusters is consistent with the classical view of bulge and halo populations of MW GCs, with more metal rich GCs segregated in the central part of the Galaxy (\citealt{2001segc.book..223H}; \citetalias{2009ApJ...694.1498M}). 

What is interesting is the relatively low scatter of the halo type II GCs in this plane if compared with the large dispersion of type I ones.
In order to better quantify this result, we considered the sample of type I GCs with [Fe/H]$<-1$ ($35$ type I GCs satisfy this condition). We obtain ${\rm r_{S}}=-0.39$ for them and ${\rm r_{S}}=0.61$ for type II GCs with [Fe/H]$<-1$ for the [Fe/H] vs Galactocentric distance relation.

The probability of obtainig a sample that displays an equal or higher $\mid {\rm r_{S}} \mid$, randomly extracting ten points from the combined samples of type I and type II GCs with [Fe/H]$\leq-1$, ($45$ GCs in total) is less than $1.5\%$. 

\begin{table}
  \caption{Oosterhoff type for type II clusters. Galactocentric distances have been also reported.
  References legend is: 1, \protect\citet{2009Ap&SS.320..261C}; 2, \protect\citet{2010AJ....139..357Z}; 3, \protect\cite{2010MNRAS.406..329S}; 4, \protect\citet{2011PASP..123.1044V}; 5, \protect\citet{2013AJ....146..119K}; 6. \protect\citet{2014MNRAS.444.1862S}.}
  \label{tab:dyn:oot}
  \centering
  \begin{tabular}{l|c c l}
              & ${\rm r}_{GC}$ (kpc) & OoT & Ref.  \\
    \hline
    NGC\,6388 &  3.1                & III & 1; 6 \\
    NGC\,6656 &  4.9                & II  & 1; 4; 5 \\
    NGC\,5139 &  6.4                & II  & 1; 4 \\
    NGC\,5286 &  8.9                & II  & 1; 2 \\
    NGC\,362  &  9.4                & I   & 1; 4 \\
    NGC\,7078 & 10.4                & II  & 1; 4 \\
    NGC\,7089 & 10.4                & II  & 1; 4 \\
    NGC\,6934 & 12.8                & I   & 1; 4 \\
    NGC\,1851 & 16.6                & I   & 1; 4 \\ 
    NGC\,1261 & 18.1                & I   & 1; 4 \\
    NGC\,6715 & 18.9                & Int & 1; 3 \\
  \end{tabular}
\end{table}

Such level of correlation might indicate the existence of a trend for the majority of type II GCs, with metal-rich ones located, nowadays, at larger ${\rm r_{GC}}$ than those with lower $[{\rm Fe}/{\rm H}]$ values. Specifically, the slope of the regression line (solid red line in Figure \ref{fig:dyn:distmetal}) is $0.033\pm0.005$. For type I GCs with $[{\rm Fe}/{\rm H}]\leq-1$, the slope of the regression line results to be, instead, $-0.037\pm0.001$. In order to further explore this possibility, we investigated the spatial distribution of the Oosterhoff type of these clusters. We recall that this quantity correlates with the metallicity of GCs \citep{1955AJ.....60..317A}. In Table \ref{tab:dyn:oot} we have listed the Oosterhoff type as well as Galactocentric distance for type II clusters. The resulting distribution seems to support the presence of such trend.

As previously done for the age-${\rm r_{GC}}$ relation, we studied the possible relations existing between metallicity and each single Galactic cartesian coordinate. The ${\rm r_{S}}$ calculated in each case are reported in Table \ref{tab:dyn:sprkc}. Type II GCs with [Fe/H]$\leq -1$ show higher values of correlation (${\rm r_{S}}=\sim 0.70$) in the X and Z coordinates than in Y (${\rm r{s}}=0.27$).

\subsection{Orbital parameters vs. cluster type}\label{sec:dyn:orb}

\begin{figure}
  \includegraphics[width=\columnwidth]{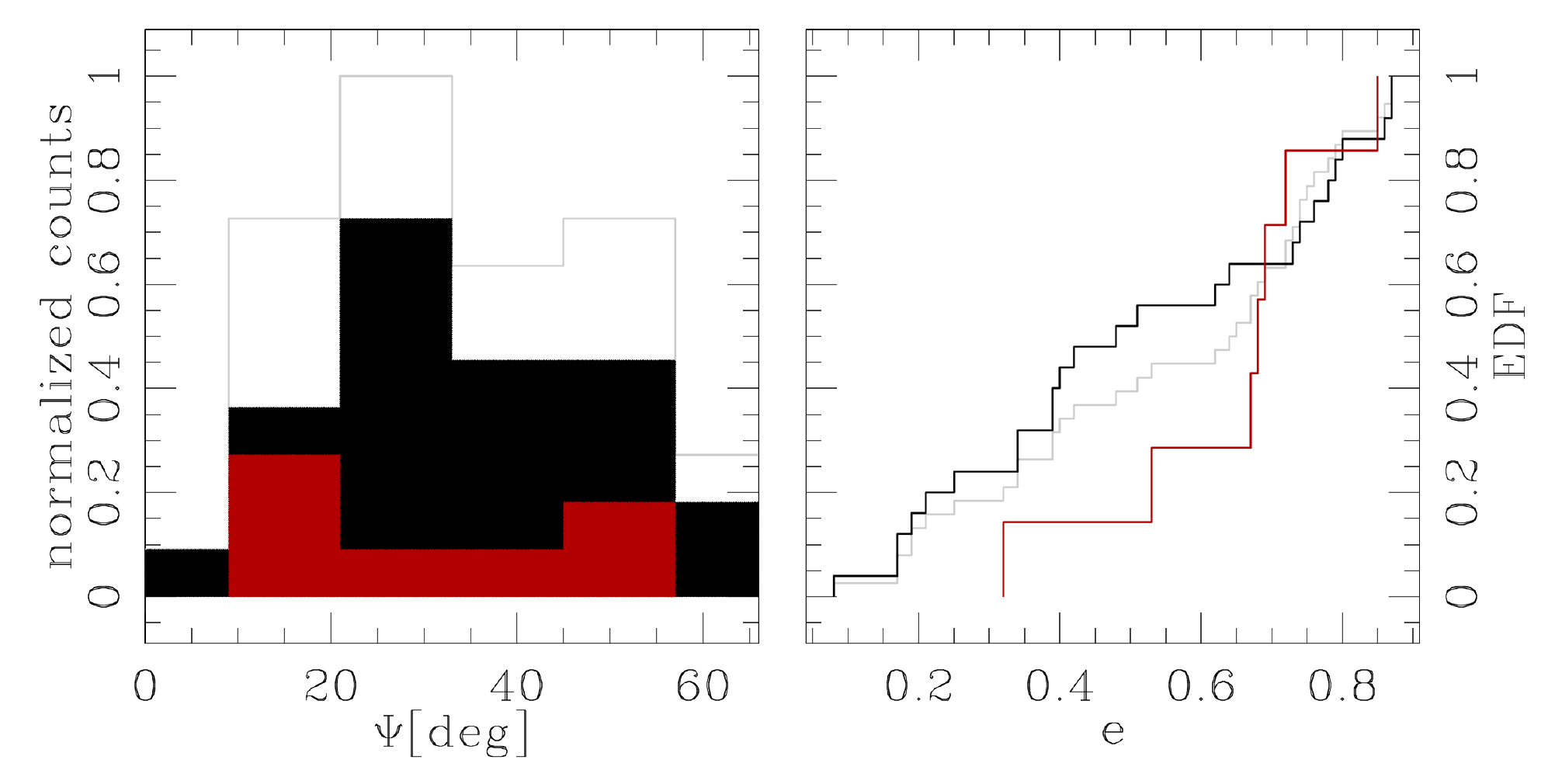}
  \caption{ Left panel: Orbit inclinations distribution. Orbit inclinations are measured with respect to the Galactic plane. The red histogram refers to type II GC; the black one to type I clusters; the gray line shows the distribution of the total sample. Right panel: Orbit eccentricity cumulative distribution. The same color-code has been adopted. The used values are from \citet{1999AJ....117.1792D}.}
  \label{fig:dyn:psiecc}
\end{figure}

Figure \ref{fig:dyn:psiecc} shows the orbit inclination distributions (left panel) and the orbital eccentricity cumulative distribution (right panel). Data are from \citet{1999AJ....117.1792D}. While the distribution of the orbit inclinations for Type II clusters (in red) is flat, these GCs appear to have preferentially high eccentricity orbits. This selection effect is particularly interesting, also in the light of the evidences provided in Section \ref{sec:dyn:pca}. The PCA, in fact, indicates a possible correlation between cluster type and orbit eccentricity.

\subsection{Cluster mass  vs. cluster type}\label{sec:dyn:mass}

PCA indicates that $16\%$ of variability in the considered sample is explained by the third principal component. This component mainly correlates with cluster mass (see lower-left panel of Figure \ref{fig:dyn:pca123}).

\begin{figure}
  \includegraphics[width=\columnwidth]{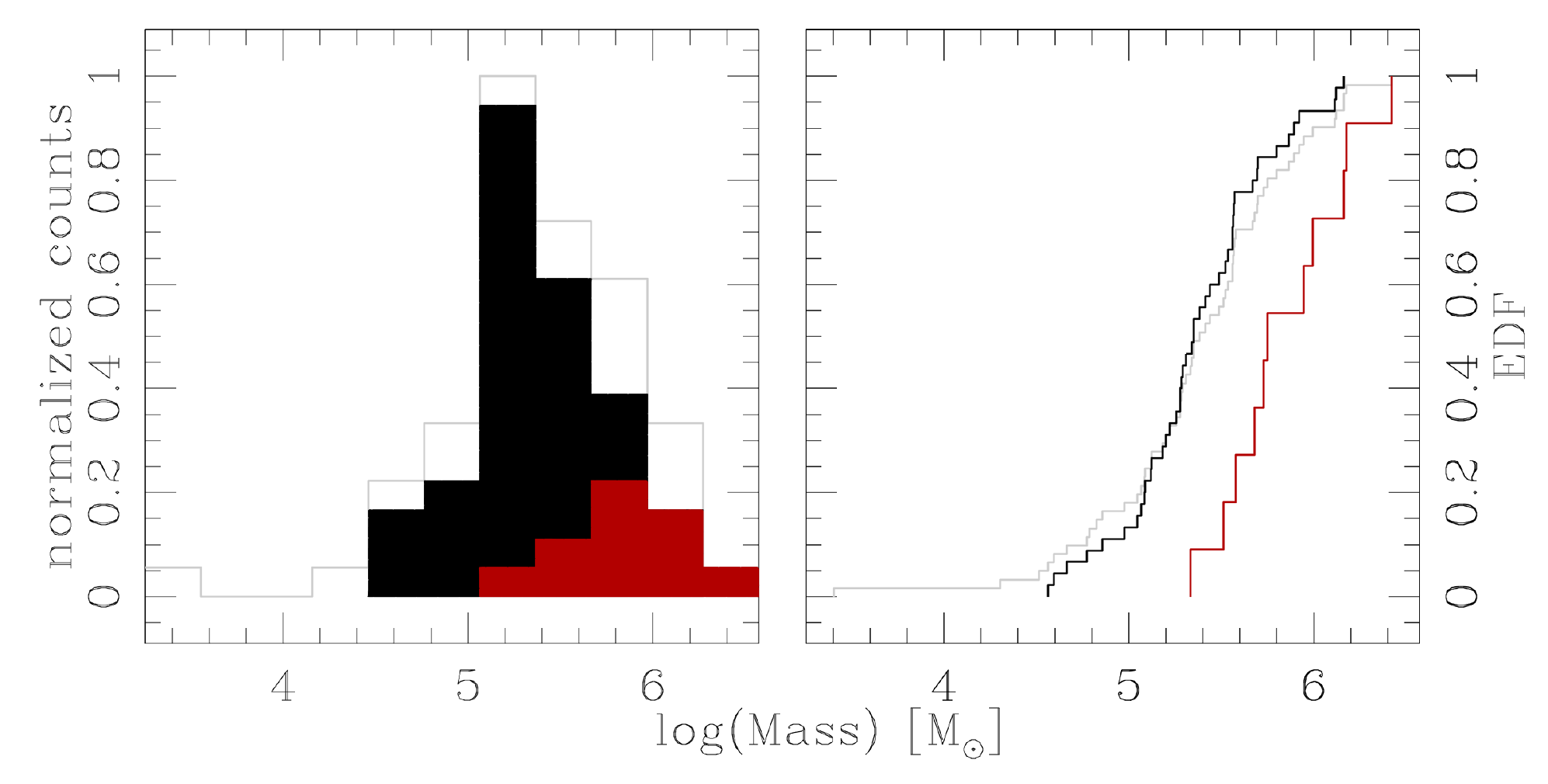}
  \caption{Left panel: Mass distribution. The red histogram refers to type II GCs; the black one to type I clusters; the gray line shows the distribution of the total sample. Right panel: Mass cumulative distribution. The same color-code has been adopted. Masses are from \protect\citet{1997ApJ...474..223G}.}
  \label{fig:dyn:massh}
\end{figure}

This is also evident in Figure \ref{fig:dyn:massh} that shows that type II clusters (in red) have, on average, higher mass than type I clusters.

\subsection{Spatial distribution of cluster types}\label{sec:dyn:spdist}

In Section \ref{sec:dyn:agemet} and Section \ref{sec:dyn:distmet} we have noted that a possible relation involving age or metallicity and the spatial distribution of type II GCs may exist. In particular, the ${\rm r_{S}}$ values indicate that both relations are somewhat stronger in Z coordinates.

\begin{figure*}
  \includegraphics[width=\textwidth]{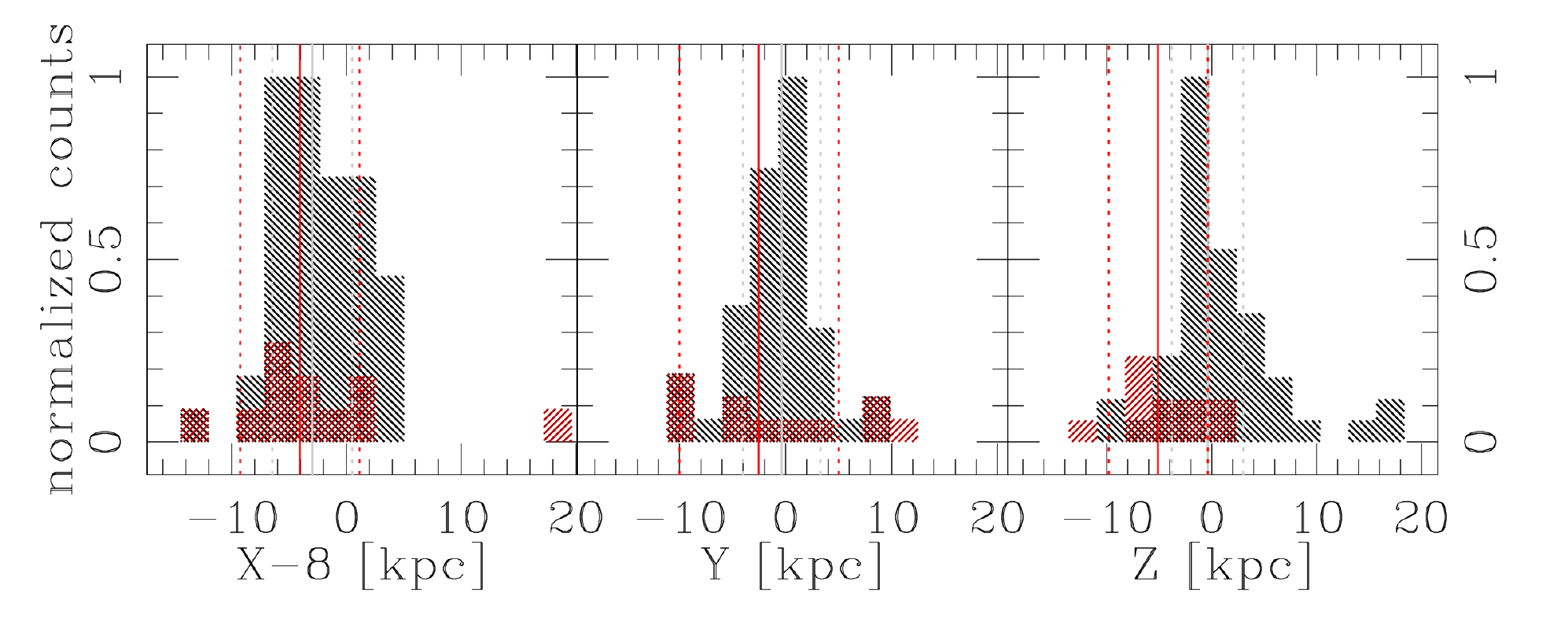}
  \caption{Spatial distribution of type I (black) and type II (red) GCs in Galactic cartesian coordinates. For each projection we have also indicated the position of the mean of each sample and the associated $68.27$ percentile deviations. Grey vertical lines refers to type I GCs, while the red ones to type II GCs.}
  \label{fig:dyn:xyzh}
\end{figure*}

The spatial distributions in the Galactic coordinates X, Y and Z are shown in Figure \ref{fig:dyn:xyzh}. The black histograms show type I GCs while the red histograms show type II ones. It can be noted that, in X and Y, the two samples seem to be equally distributed around $0$. In the Z coordinate, instead, type II GCs seem to be preferentially located at negative values, in contrast with the apparent symmetry of the type I distribution. Specifically, no type II GCs are found above Z$\sim 2$. 
Calculating the median of each sample and the associated dispersion, we measured the separation between the two distributions in each projection. It results that, in units of $\sigma$, the two medians are separated by $0.17$ in the X coordinate, $0.26$ in Y and $0.81$ in Z. We thus calculated the probability of obtaining such values of separation between medians, randomly extracting $11$ points from the joint sample of type I and type II GCs. The resulting probabilities are $\sim 65\%$ for X, $\sim30\%$ for Y and $<0.3\%$ for Z, which points to a likely bias in the Z values of the type II sample.

\section{Summary and Conclusion}
Chromosome maps (\citealt{2015ApJ...808...51M}, \citetalias{2017MNRAS.464.3636M}) showed that virtually all Galactic GCs host multiple stellar populations. Chromosome maps also allowed us to identify a family of Galactic GCs that shows an above-the-mean complexity in their stellar population. These clusters were classified as type II clusters, the remaining ones being labelled as type I clusters \citepalias{2017MNRAS.464.3636M}.

In this paper we study if and how cluster type correlates with GC properties. We performed principal component analysis based on 11 quantities: cluster type, Oosterhoff type, metallicity, relative age, orbit total energy, orbit total angular momentum, orbit eccentricity, orbit inclination with respect to the Glactic plane, cluster masses, core radii and tidal radii. The main sources of variance in the considered sample results to be contained in orbital parameters and, at a lesser extent, in the relation between cluster type and relative age.
Below we summarize the main results.
\renewcommand{\labelitemi}{$-$}
\begin{itemize}
  \item In the plane age-metallicity, type II clusters define a clear trend, with more metal-rich clusters being younger. NGC\,6388 is an exception, but it also is the only type II cluster located in the bulge.
  \item There are hints of a possible relation between age and Galactocentric distance for type II GCs, with younger clusters being located outwards.
  \item In the [Fe/H]-${\rm r_{GC}}$ plane, halo type II clusters show a trend, with more metal-rich clusters being located outwards. This trend is consistent with the radial distribution of their Oosterhoff type and apparently inconsistent with the trend found for type I GCs.
  \item The orbits of type II clusters are more eccentric than the average of type I clusters. A large dispersion of the orbital inclinations with respect to the Galactic plane is also observed and is similar for both groups.
  \item Type II clusters are, on average, more massive than type I ones. There are, in any case, type I clusters with mass comparable to those of type II.
  \item Nowadays, type II GCs are preferentially located below the plane of the Galaxy. Specifically, no type II GC has been observed above Z$\sim2$ kpc.  
\end{itemize}

This suggest two further investigations: (i) it will be of primary interest to search for additional type II clusters, especially at Z$>0$ kpc; (ii) it would be quite informative to perform detailed simulation of the kinematics of type II GCs. 

For the first point, we are suggesting to trace the path to a photometric survey aimed at enlarging the sample of Galactic GCs for which chromosome maps could be used to discern their type. Extending the survey to the full sample of the MW will improve the statistical significance of the present results.

For the second point we find of particular interest to establish, in the light of this new classification, if some remnants of a common dynamical pattern could be detected among the known type II clusters. In fact, the high incidence of negative values of Z for type II GCs could indicate a common extragalactic origin for these systems.

\section*{Acknowledgements}
We thank the anonymous referee for the careful revision that improved the quality of the present manuscript. M.S., A.A. and G.P. acknowledge support from the Spanish Ministry of Economy and Competitiveness (MINECO) under grants AYA2013-42781 and AYA2017-89841-P. M.S. and A.A. acknowledge support from the Instituto de Astrof{\'i}sica de Canarias (IAC) under grant 309403. G.P. acknowledges partial support by the Universit{\`a} degli Studi di Padova Progetto di Ateneo CPDA141214 and by INAF under the program PRIN-INAF2014.

\bibliographystyle{}

\bsp	
\label{lastpage}
\end{document}